\documentclass[prb,showpacs,superscriptaddress,endfloats]{revtex4}
\usepackage{graphics} 
\usepackage{epsfig} 

\begin{document}   
\title{Lifetime of d-holes at Cu surfaces: Theory and experiment} 

\author{A. Gerlach} 
\author{K. Berge} 
\author{A. Goldmann}

\affiliation{Fachbereich Physik, Universit\"at Kassel,
  Heinrich-Plett-Strasse 40, D--34132 Kassel, Germany}

\author{I. Campillo}

\affiliation{Materia Kondentsatuaren Fisika Saila, Euskal Herriko
  Unibertsitatea, 644 Posta Kutxatila, 48080 Bilbo, Basque Country,
  Spain}

\author{A. Rubio}

\affiliation{Materialen Fisika Saila, Kimika Fakultatea, Euskal
  Herriko Unibertsitatea and Centro Mixto CSIC-UPV/EHU, 1072 Posta
  kutxatila, 20080 Donostia, Basque Country, Spain}

\affiliation{Donostia International Physics Center (DIPC), Paseo
  Manuel de Lardizabal s/n 20018, Donostia, Basque Country, Spain}

\author{J. M. Pitarke} 

\affiliation{Materia Kondentsatuaren Fisika Saila, Euskal Herriko
  Unibertsitatea, 644 Posta Kutxatila, 48080 Bilbo, Basque Country,
  Spain}

\affiliation{Donostia International Physics Center (DIPC), Paseo
  Manuel de Lardizabal s/n 20018, Donostia, Basque Country, Spain}

\author{P. M. Echenique}

\affiliation{Materialen Fisika Saila, Kimika Fakultatea, Euskal
  Herriko Unibertsitatea and Centro Mixto CSIC-UPV/EHU, 1072 Posta
  kutxatila, 20080 Donostia, Basque Country, Spain}

\affiliation{Donostia International Physics Center (DIPC), Paseo
  Manuel de Lardizabal s/n 20018, Donostia, Basque Country, Spain}

\date{\today}

\begin{abstract}  
  We have investigated the hole dynamics at copper surfaces by
  high-resolution angle-resolved photoemission experiments and
  many-body quasiparticle GW calculations. Large deviations from a
  free-electron-like picture are observed both in the magnitude and
  the energy dependence of the lifetimes, with a clear indication that
  holes exhibit longer lifetimes than electrons with the same
  excitation energy. Our calculations show that the small overlap of
  $d$- and $sp$-states below the Fermi level is responsible for the
  observed enhancement. Although there is qualitative good agreement
  of our theoretical predictions and the measured lifetimes, there
  still exist some discrepancies pointing to the need of a better
  description of the actual band structure of the solid.
\end{abstract}

\maketitle
\section{Introduction}  
Electron and hole dynamics in the bulk as well as at the various
surfaces of a solid play a key role in a great number of physical and
chemical phenomena\cite{dai95,petek97,petek00}. These range from
photon-induced surface reactions at metals to device-physical
applications in semiconducting nanostructures. It is evident that both
reliable experimental information and detailed theoretical
understanding are necessary prerequisites for any future
application-oriented tailoring of dynamical electronic properties.
Seen from theory any excitation of an electron-hole quasiparticle
reflects basic properties of the solid as a correlated many-electron
system, which must necessarily be described with inclusion of the very
details of band structure, symmetry and dimensionality, and also the
various excitation and decay channels.  From the experimental point of
view, additional contributions to experimental spectra result from
effects like coupling of electrons and holes to phonons and their
scattering with structural defects, and it is a sometimes tedious task
to distinguish these effects
unambiguously\cite{matzdorf98,theilmann97}.

At present our quantitative understanding of the excitation mechanisms
and their relaxation channels is far from being complete: There is
ample experimental evidence from experiments at copper
surfaces\cite{pawlik97,knoesel98,petek00.2} that hot-electron
lifetimes $\tau$ probed by time-resolved two-photon photoemission
(TR-TPPE) techniques do not follow an inverse quadratic dependence
$\tau\propto(E-E_F)^{-2}$, which would result from simple phase-space
considerations\cite{echenique00}. In contrast, an analysis of the
standing-wave patterns observed by scanning tunneling spectroscopy on
Cu(111) has been shown to yield an accurate quadratic behaviour in the
same energy range, i.e.\ up to about $3\, \mathrm{eV}$ above the Fermi
level $E_F$ \cite{buergi99}. In this tunneling experiment electrons
are injected into a well-defined surface state band, and the coherence
length $l$ is then derived from the spatial damping of the
standing-wave pattern produced by repeated reflections between
parallel steps. Simple and reliable model assumptions connect $l$ with
$\tau$ and allow the determination of $\tau(E-E_F)$.  TR-TPPE,
however, uses femtosecond pump-probe techniques to monitor the
time-dependent occupation of the intermediate energy levels. These may
have decay and excitation channels which differ from those of the
tunneling experiment. One aspect appears particularly important:
Photons with typical energies of $3 - 4 \, \mathrm{eV}$ do not only
probe $sp$-like bulk and surface electrons, but in some cases also
electrons from occupied $d$-bands are
excited\cite{pawlik97,knoesel98,petek00.2,matzdorf99}.  Regarding the
above mentioned TR-TPPE-experiments on copper, where the upper edge of
the $d$-bands is located $2.0\, \mathrm{eV}$ below the Fermi level, a
still ongoing debate considers whether\cite{knoesel98} or
not\cite{petek00.2} Auger recombination of long-lived $d$-holes
contributes significantly to the hot-electron generation.

Since neither the magnitude nor the energy dependence of $d$-hole
lifetimes is precisely known, we consider this subject to be of
significant current interest.  The new experimental data for copper
presented in this paper were obtained using one-photon photoemission
spectroscopy\cite{kevan92,huefner95}.  This technique allows to
determine lifetimes only from a detailed analysis of
linewidths\cite{smith93,mcdougall95}, and requires simultaneously low
temperatures, excellent energy as well as angular resolution and in
general tunable photon energy. This expense is more than compensated
by the very detailed understanding of the one-photon excitation
channel and the possibility to locate a photo-hole exactly with
respect to both initial-state energy $E_i$ and wave vector
$\mathbf{k}$. Recent calculations have shown that any reliable theory
of hot-electron and hole lifetimes in metals must go beyond a
free-electron description of the solid
\cite{campillo00,schoene00,campillopress}.  In Sec.\ III of this paper
we compare many-body quasiparticle GW calculations of $d$-hole
lifetimes in copper, which have already been presented in Ref.\ 
\onlinecite{campillopress}, to our photoemission data.

\section{Experimental results} 

\subsection{Experimental setup} 

Our experiments with synchrotron radiation were performed at the
storage ring Bessy~I in Berlin. Normal emission photoelectron spectra
were collected using the 2m-Seya beamline and a high-resolution
photoemission station described and characterized in detail in
Ref.~\onlinecite{janowitz99}.  The sample is mounted on a manipulator
cryostat having five degrees of freedom: $x,y,z$ translation, rotation
around the manipulator axis and rotation around the surface normal,
the latter one being particularly useful for polarization dependent
suppression of individual photoemission lines.  The sample temperature
could be varied between $20\, \mathrm{K}$ and room temperature. A
sample load lock system with a transfer rod allows to decouple the
sample from the manipulator for preparation by argon ion bombardment
and annealing. The energy resolution including the photon
monochromator was set to $30 \, \mathrm{meV}$, as verified by the
analysis of the Fermi edge emission taken at $T=22\, \mathrm{K}$.  The
angular resolution of $\Delta\theta=\pm 1^{\circ}$ is sufficient for
normal emission spectra taken at flat bands.  The experiments with
Cu(100) were performed using a Scienta high-resolution photoelectron
spectrometer in our home laboratory. It is equipped with facilities
for LEED, monochromatized XPS and UPS, the standard methods for sample
cleaning and annealing, and a closed-cycle refrigerator for sample
cooling. The spectra reported below were excited using monochromatized
HeII radiation ($\hbar\omega = 40.8 \, \mathrm{eV}$) from a Gammadata
microwave light source. The angular resolution was estimated to be
about $\pm 0.5^{\circ}$; the energy resolution was better than $8\,
\mathrm{meV}$.

The copper samples were oriented to better than $\pm 0.25^{\circ}$,
polished first mechanically and then electrochemically.  After
insertion into the UHV chamber they were cleaned and prepared by
repeated cycles of argon ion bombardment and an extended annealing
procedure at $T = 700 \, \mathrm{K}$.  The high surface quality had
been verified experimentally: In these measurements we obtained very
sharp LEED spot profiles of about 1\% of the surface Brillouin zone
diameter and, even more significant, sharp photoemission peaks: The
Shockley-type surface state on Cu(110), residing at $E_i=-0.43 \,
\mathrm{eV}$ around the $\overline{Y}$-point\cite{kevan92,huefner95},
shows a linewidth of less than $80 \, \mathrm{meV}$ (FWHM) at room
temperature.  The width of the $d$-like Tamm surface state on Cu(100),
residing at $E_i=-1.80 \, \mathrm{eV}$ around the $\overline{M}$-point
of the corresponding surface Brillouin zone\cite{theilmann97}, is $35
\, \mathrm{meV}$ at room temperature.

\subsection{Linewidth information}

\begin{figure}
  \begin{center}
    \includegraphics[width=14cm]{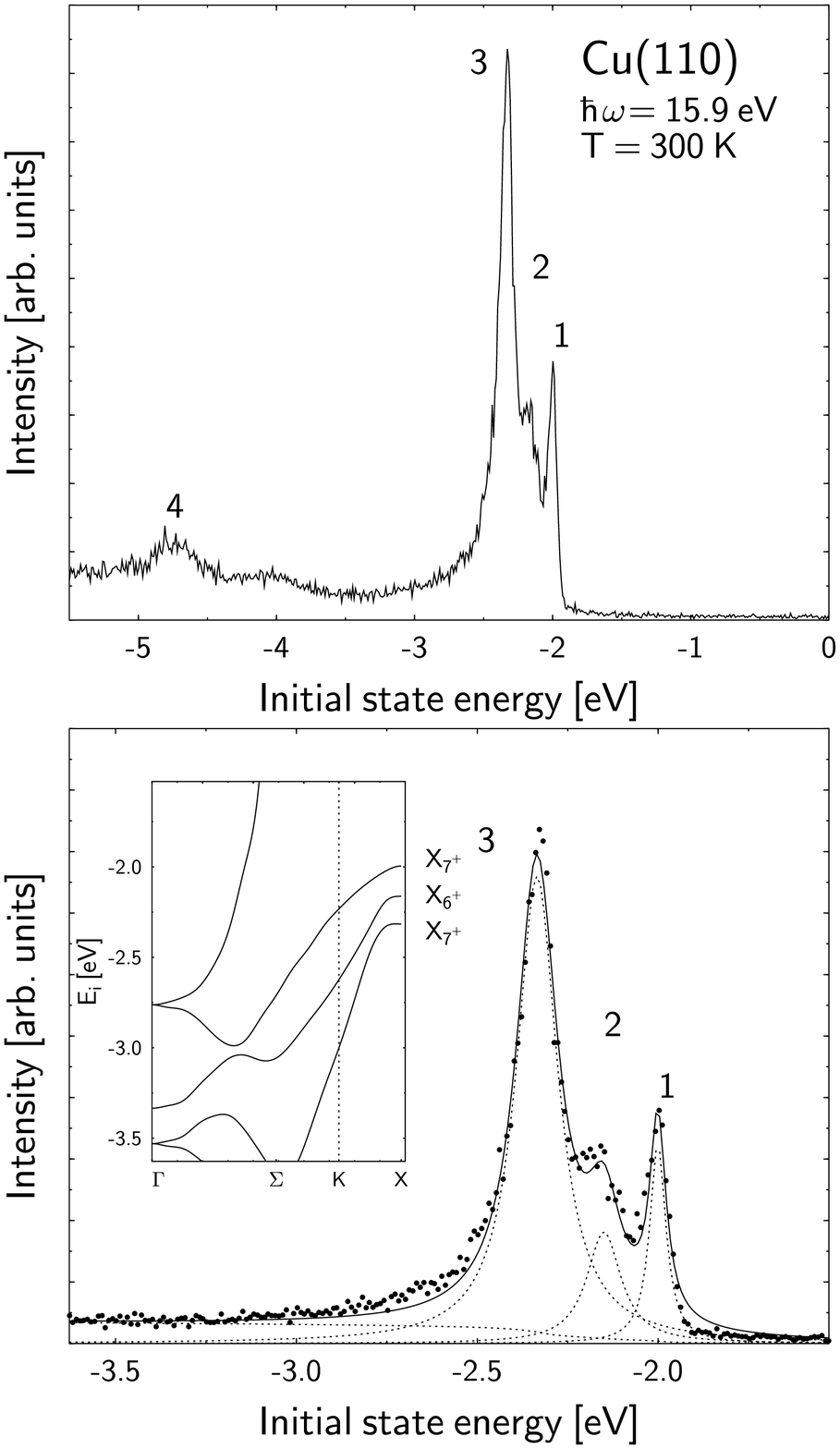}    
    \caption{Top: Normal emission photoelectron spectrum, as obtained at
      room temperature from Cu(110) with photons of energy
      $\hbar\omega=15.9 \, \mathrm{eV}$. Peak numbers refer to the
      data collected in table~\ref{tab:linewidth}.  Bottom:
      decomposition of peaks 1 to 3 using three Lorentzians and a
      Shirley-type background. The inset shows part of the copper band
      structure from Ref.~\protect\onlinecite{eckardt84}\protect{} along
      the $\Sigma$-direction ($\Gamma$KX).}
    \label{fig:cu159edc}
  \end{center}
\end{figure}

An experimental photoemission linewidth $\Gamma_{exp}$ generally has
contributions from both the electron and the hole lifetime. This is
discussed in great detail in Ref.~\onlinecite{smith93} (see also
Refs.~\onlinecite{matzdorf98,kevan92}, and~\onlinecite{huefner95}).
For normal emission, the measured linewidth $\Gamma_{exp}$ is
approximately given by
\begin{equation}
  \label{eq:smith} 
  \Gamma_{exp}=\left(\Gamma_h+\frac{v_h}{v_e}\Gamma_e\right) 
  \left(\left|1-\frac{v_h}{v_e}\right|\right)^{-1},
\end{equation}  
where $\Gamma_h = \hbar / \tau_h, \Gamma_e = \hbar / \tau_e$ are the
final-state hole and electron state inverse lifetimes, and $v_h, v_e$
are the corresponding group velocities $v = \hbar^{-1}\mid\partial
E/\partial k\mid$ taken normal to the surface.  Although the copper
$d$-bands show a weak dispersion, i.e.\ $v_e \gg v_h$ (see e.g.\ 
Refs.~\onlinecite{matzdorf98} and \onlinecite{courths84}), the
measured linewidth $\Gamma_{exp}$ is generally dominated by the
contribution of the photoelectron, because $\Gamma_e$ exceeds
$\Gamma_h$ by more than an order of magnitude.  In order to obtain
$\Gamma_h$, we have to tune the photon energy to direct transitions
with $v_h=0$, i.e.\ to symmetry points of the band structure along the
surface normal. In the experiments with Cu(110) we observed
transitions around the X-point.  At the upper $d$-band edge the bands
are split under the combined action of crystal-field and spin-orbit
coupling: Three subbands labeled X$_{7^+}$ ($E_i=-2.00 \,
\mathrm{eV}$), X$_{6^+}$ ($-2.15 \, \mathrm{eV}$), and X$_{7^+}$
($-2.34 \, \mathrm{eV}$) in relativistic double-group symmetry
notation are observed\cite{huefner95,courths84}, see inset in
Fig.~\ref{fig:cu159edc}.  Also at X$_{7^+}$ ($E_i=-4.80 \,
\mathrm{eV}$) a $d$-hole may be created.  Direct transitions at the
$\Gamma$-point may be excited using photons of $\hbar\omega = 40.8 \,
\mathrm{eV}$ (HeII radiation) at the Cu(100) surface\cite{courths84}.
Thus this experiment probes states at $\Gamma_{8^+}$ ($E_i=-3.59 \,
\mathrm{eV}$), $\Gamma_{7^+}$ ($-3.40 \, \mathrm{eV}$) , and
$\Gamma_{8^+}$ ($-2.82 \, \mathrm{eV}$) . In summary, we have access
to several hole states within the $d$-bands. A determination of their
lifetimes by linewidth analysis should be sufficient to recognize
trends in $\tau_h(E_i)$.

We collected normal emission spectra from Cu(110) using photon
energies between $14$ and $21 \, \mathrm{eV}$, both at room
temperature and $T=25\, \mathrm{K}$.  A typical spectrum taken at
$\hbar\omega=15.9 \, \mathrm{eV}$ and $T=300 \, \mathrm{K}$ is
reproduced in Fig.~\ref{fig:cu159edc}(a).
Four peaks, which correspond to transistions from $\mathbf{k}$-points
near X$_{7^+}$, X$_{6^+}$, X$_{7^+}$, and X$_{7^+}$, are clearly
resolved.  Assuming a Shirley-type background\cite{shirley72} to
account for the underlying continuous emission and Lorentzian line
shapes, all spectra were decomposed in order to obtain the initial
state energy $E_i$, intensity, and linewidth $\Gamma_{exp}$ in their
dependence on $\hbar\omega$.  This kind of analysis, as shown in
Fig.~\ref{fig:cu159edc}(b), works quite well. But especially peak 1
shows an asymmetric lineshape with a broader tail to higher binding
energies, which cannot be reproduced by our fit.  A detailed line
shape analysis, based on a suitable parameterization of $\Gamma_h$ and
$\Gamma_e$, reveals that this discrepancy is related to the decreasing
inverse lifetime $\Gamma_h$ at the upper d-band edge and the
uncertainty of $k_\bot$ along the $\Sigma$-direction ($\Gamma$KX), see
further below.

\begin{figure}
  \begin{center}
    \includegraphics[width=14cm]{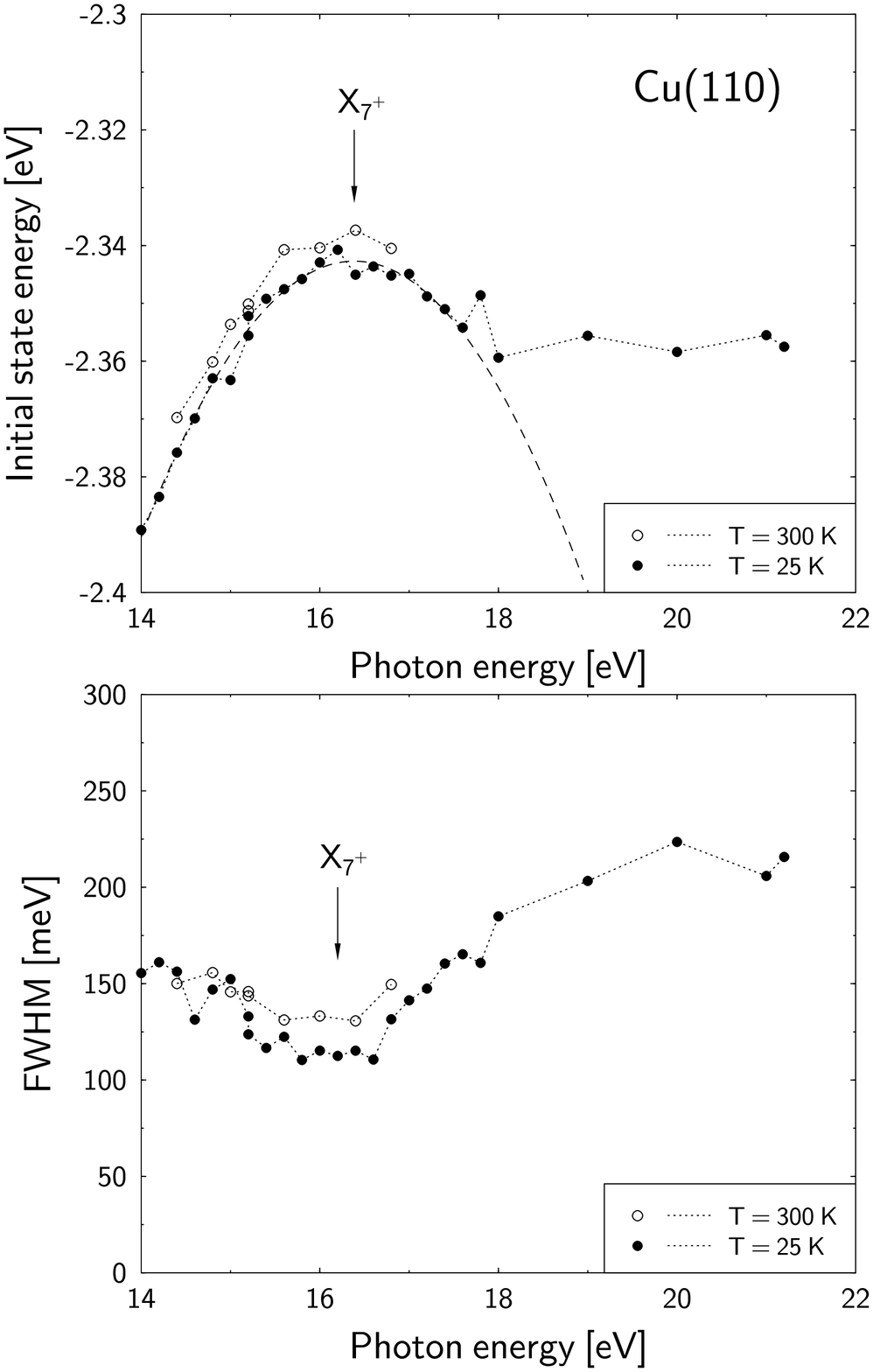}     
    \caption{Top: Energy position of peak 3 in 
      Fig.~\ref{fig:cu159edc}, as a function of the photon energy
      $\hbar\omega$. Note the asymmetric dispersion around $\hbar
      \omega = 16.2 \, \mathrm{eV}$. Bottom: Full width at half
      maximum of the adjusted Lorentzian line.  The minimal linewidth
      corresponds to transitions at the X-point with $v_h = 0$.}
    \label{fig:cudisp}
  \end{center}
\end{figure}

\begin{figure}
  \begin{center}
    \includegraphics[width=14cm]{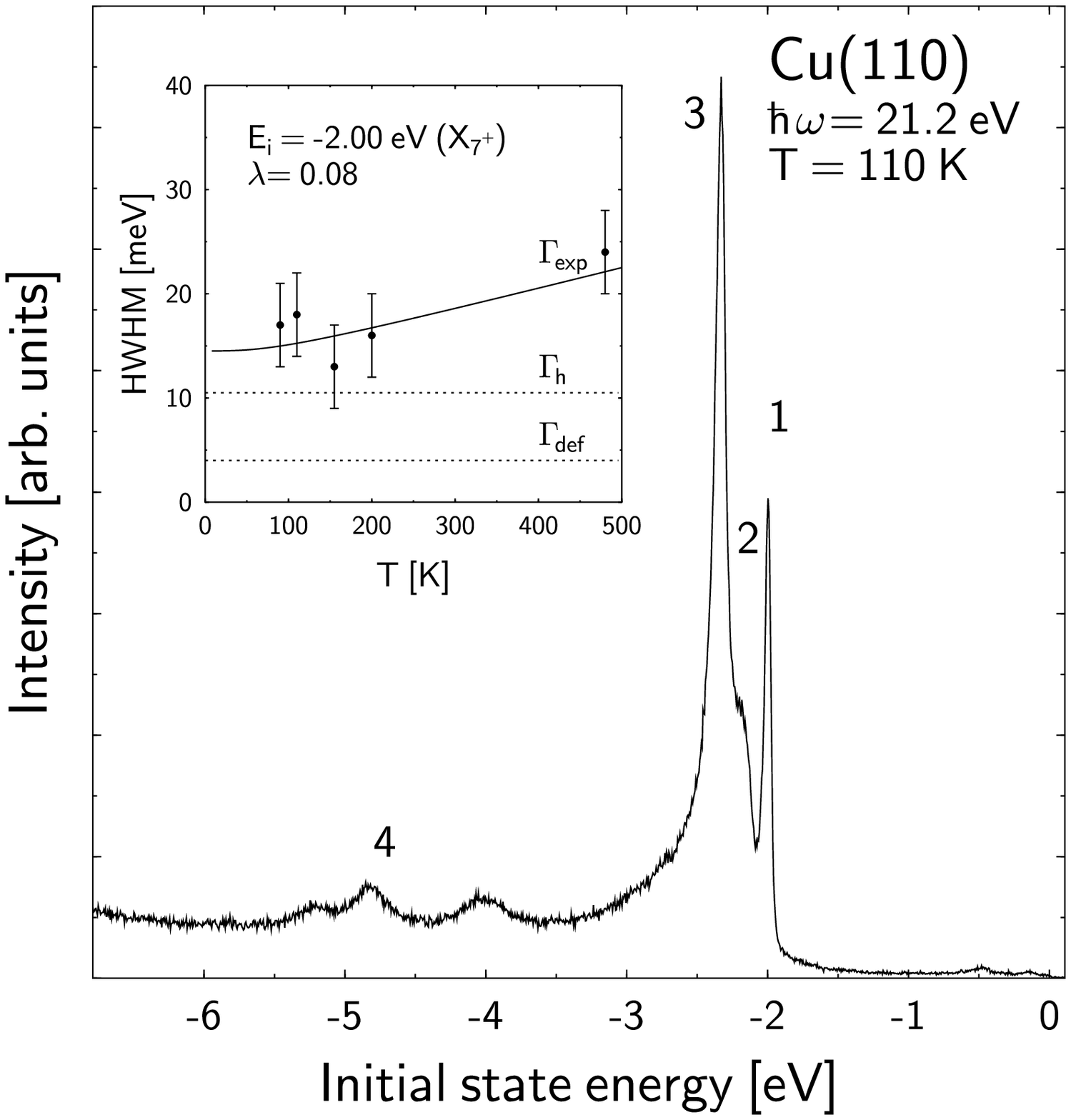}
    \caption{Cu(110) normal emission spectrum obtained with $21.2 \, 
      \mathrm{eV}$, Inset: Temperature dependent contributions to the
      half-width at half-maximum of peak 1. For details on phonon and
      defect scattering see text.}
    \label{fig:cu212edc}
  \end{center}
\end{figure}

Fig.~\ref{fig:cudisp} shows results for transitions out of the third
band below $E_i = -2.34 \, \mathrm{eV}$.  The observed peak shift with
$\hbar \omega$ is not symmetric around $\hbar \omega = 16.2 \,
\mathrm{eV}$, as might be expected from the symmetry along the
$\Sigma$-direction in the Brillouin zone. We attribute this to the
presence of the well-known band gap above the Fermi
level\cite{dietz79,courths84}: Photoelectrons excited with
$\hbar\omega=17 - 21 \, \mathrm{eV}$ cannot couple to the
corresponding bulk states.  In consequence, only ``surface
emission''\cite{huefner95,courths84} into evanescent final states
contributes to the photoemission intensity, which drops significantly
within this photon-energy region.  As is evident from
Fig.~\ref{fig:cudisp}(b), the minimum linewidth is obtained at $E_i =
-2.34 \, \mathrm{eV}$ around $\hbar\omega=16.2\, \mathrm{eV}$, i.e.\ 
just at the X$_{7^+}$-point.  This is consistent with
eq.~(\ref{eq:smith}).  The linewidth $\Gamma_{exp}$ gets smaller by
approaching the X-point with $v_h = 0$. Similar data were obtained for
the other peaks, of course with the maxima and minima located at
different photon energies.
In case of peak 1, the experimental resolution exceeds the
lifetime-width, and our data analysis gets uncertain. Therefore we
have taken additional data in our home laboratory using the HeI
radiation at $\hbar\omega = 21.2 \, \mathrm{eV}$, which allowed to
determine the half-width at half-maximum (HWHM) of peak 1 at its lower
binding energy side with higher precision. A typical spectrum recorded
at $T = 110 \, \mathrm{K}$ is reproduced in Fig.~\ref{fig:cu212edc}.
Although the excitation is not exactly related to the X-point, the
aforementioned band gap ``pins'' the transition right next to the
symmetry point\cite{huefner95,courths84}.  We regard the linewidth of
peak 1 obtained with $\hbar\omega = 21.2 \, \mathrm{eV}$ as upper
limit for $\Gamma_h$ with only a small contribution due to the
non-vanishing band dispersion.

\begin{figure}
  \begin{center}
    \includegraphics[width=14cm]{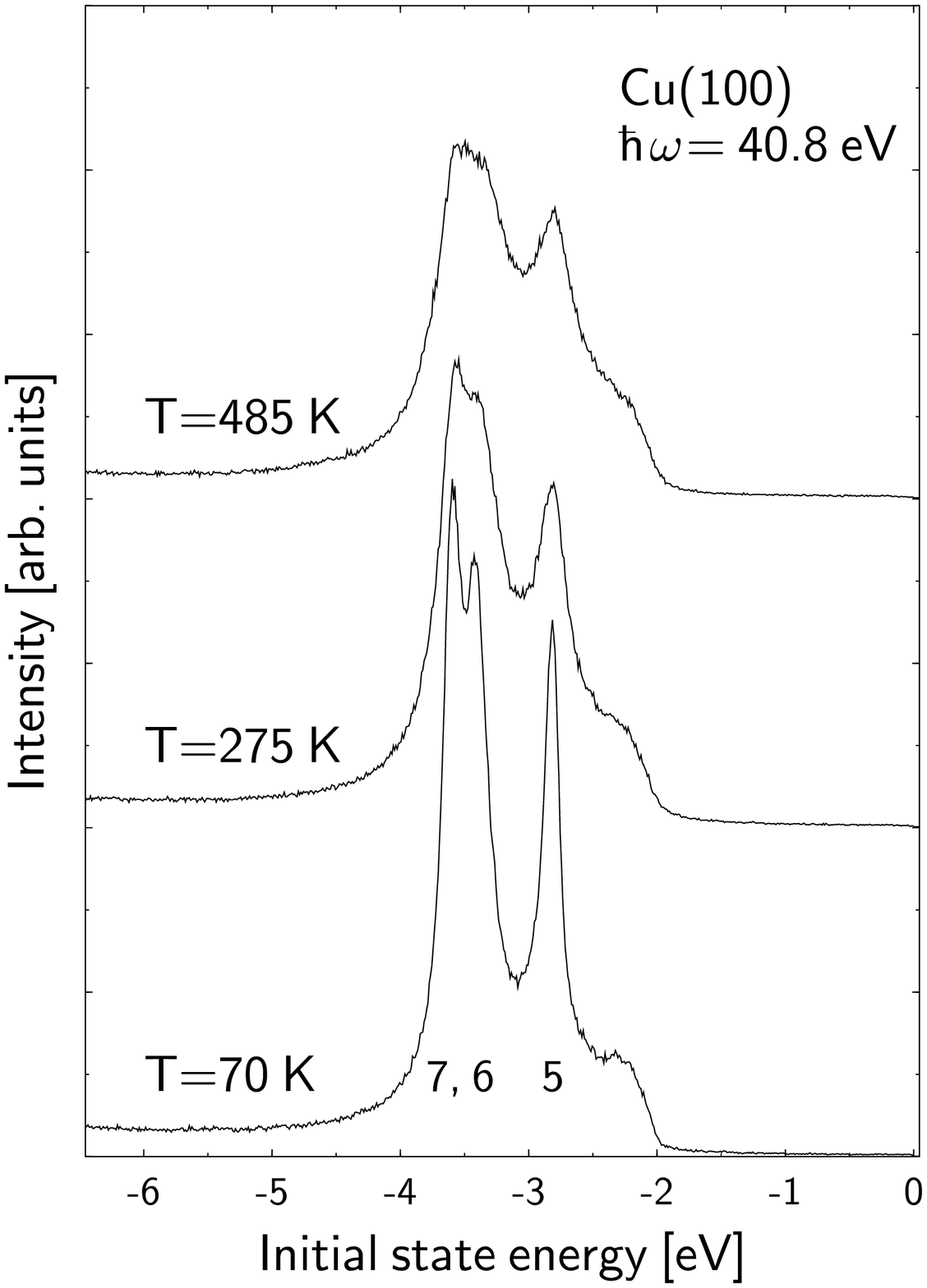}
    \caption{Normal emission photoelectron spectra, as obtained at different 
      temperatures from Cu(100) with photons of energy
      $\hbar\omega=40.8 \, \mathrm{eV}$.  Peak numbers refer to data
      collected in table~\ref{tab:linewidth}.}
    \label{fig:cuj1014edc}
  \end{center}
\end{figure}

All linewidths were extrapolated to $T = 0 \, \mathrm{K}$ to correct
for phonon-induced linewidth contributions $\Gamma_{h-ph}(T)$. Using
the well-known formula\cite{mcdougall95}
\begin{equation}
  \label{eq:elphww1}
  \Gamma_{h-ph}(T) = 2 \pi \hbar \lambda \int_{0}^{\omega_D} 
  d \omega' (\omega'/\omega_D)^2 \left( 1 - f(\omega - \omega') + 
    2n(\omega') + f(\omega + \omega') \right) \, ,
\end{equation}
where $f(\omega)$ is the Fermi-distribution, $n(\omega)$ the
Bose-distribution and $\hbar \omega_D$ the maximum phonon energy in
the Debye approximation, we extracted the mass-enhancement parameter
$\lambda$ for several photo-holes within the $d$-bands. The inset in
Fig.~\ref{fig:cu212edc} shows the temperature dependent phonon
contribution to the linewidth of peak 1 at $E_i = -2.00 \,
\mathrm{eV}$.  In this case we derive $\Gamma_{h-ph}(300 \,
\mathrm{K}) = 8 \, \mathrm{meV}$ using eq.~(\ref{eq:elphww1}) with the
experimentally determined value of $\lambda = 0.08$. Thus the
extrapolation to $T = 0 \, \mathrm{K}$ gives a phonon-corrected
linewidth of $29 \, \mathrm{meV}$. Typical spectra of Cu(100) with
$\hbar\omega=40.8 \, \mathrm{eV}$ taken at three different
temperatures are shown in Fig.~\ref{fig:cuj1014edc}.  Being somewhat
better resolved than those presented earlier\cite{matzdorf94}, they
demonstrate the dramatic influence of the temperature-dependent
broadening due to hole-phonon coupling at the $\Gamma$-point.  In
practice we never observed any deviation from the linear high
temperature limit\cite{mcdougall95,matzdorf96} $\Gamma_{h-ph} = 2 \pi
\lambda k_B T$, but there is a significant decrease of $\lambda(E_i)$
on approaching the upper $d$-band edge. We attribute this to the
reduced phase space for hole-phonon scattering in this energy range:
Since there are no $d$-states available above $E_i = -2.00 \,
\mathrm{eV}$, the phonon assisted $d$-hole decay is strongly
suppressed.

Although we use excellent quality surfaces as verified by the narrow
emission lines from the surface states, broadening from defect
scattering has to be considered\cite{theilmann97,kevan83,tersoff83}.
Recently we investigated the magnitude $\Gamma_{def}$ of such
contributions to the photoemission linewidths of various surface
states on copper\cite{theilmann97}. Our main message was that
$\Gamma_{def}$ scales inversely proportional to the effective band
mass. Using the appropriate values around the X-point as taken from
the band structure\cite{eckardt84}, we can safely estimate that
$\Gamma_{def} < 10 \, \mathrm{meV}$ for peaks $2 - 7$. Moreover, we
determined the photoemission linewidth of peak 1 for differently
annealed (and differently ordered) surfaces after argon ion
bombardment at low temperatures.  Simultaneously the width $w$ of LEED
spot-profiles is measured with high resolution. A linear correlation
of both linewidths is observed. In the present case the extrapolation
to a ``defect-free'' sample ($w \rightarrow 0$) results in a
contribution of $\Gamma_{def} = 8 \, \mathrm{meV}$.  Since the
different contributions to $\Gamma_{exp}$ add linearly
\cite{mcdougall95}, we finally get $\Gamma_h = (21 \pm 5) \,
\mathrm{meV}$. Our linewidth data are compiled in
table~\ref{tab:linewidth}.  Each value is obtained after extrapolation
to $T = 0 \, \mathrm{K}$ and treating the defect scattering as
described above.
\begin{table}[htbp]
  \begin{center}
    \begin{tabular}{c||cccr@{$\, \pm \, $}lcl}
      \hline
      Peak & Symmetry & $E_i [\mathrm{eV}]$ & sample & 
      \multicolumn{2}{c}{$\Gamma_h [\mathrm{meV}]$} & $\lambda$ & 
      Reference \\ \hline
      &X$_{7^+}$& -2.02 & Cu(100)    & 25 & 10 & -    & 
      Ref.~\onlinecite{purdie98}\\
      &X$_{7^+}$& -1.98 & Cu(100)/Cs & 28 & 3  & 0.20 & 
      Ref.~\onlinecite{petek99}\\
      1 &X$_{7^+}$& -2.00 & Cu(110) & 21  & 5  & 0.08 & this work\\
      2 &X$_{6^+}$& -2.15 & Cu(110) & 80  & 20 & -    & this work\\
      3 &X$_{7^+}$& -2.34 & Cu(110) & 102 & 20 & 0.16 & this work\\
      4 &X$_{7^+}$& -4.80 & Cu(110) & 232 & 30 & -    & this work\\
      5 &$\Gamma_{8^+}$& -2.82 & Cu(100) & 132 & 20 & 0.43 & this work\\
      6 &$\Gamma_{7^+}$& -3.40 & Cu(100) & 177 & 30 & -    & this work\\
      7 &$\Gamma_{8^+}$& -3.59 & Cu(100) & 152 & 20 & 0.54 & this work\\
      \hline
    \end{tabular}
    \caption{Upper limits for the $d$-hole inelastic linewidth 
      $\Gamma_h$ due to electron-hole interaction at various symmetry
      points of the copper bulk band structure. Also reproduced are
      data obtained at X$_{7^+}$ by Purdie et al.\ \cite{purdie98}
      using high-resolution photoemission from Cu(100) and by Petek et
      al.\ \cite{petek99} from TR-TPPE measurements, which are both in
      excellent agreement with our result.}
    \label{tab:linewidth}
  \end{center}
\end{table}

\begin{figure}
  \begin{center}
    \includegraphics[width=14cm]{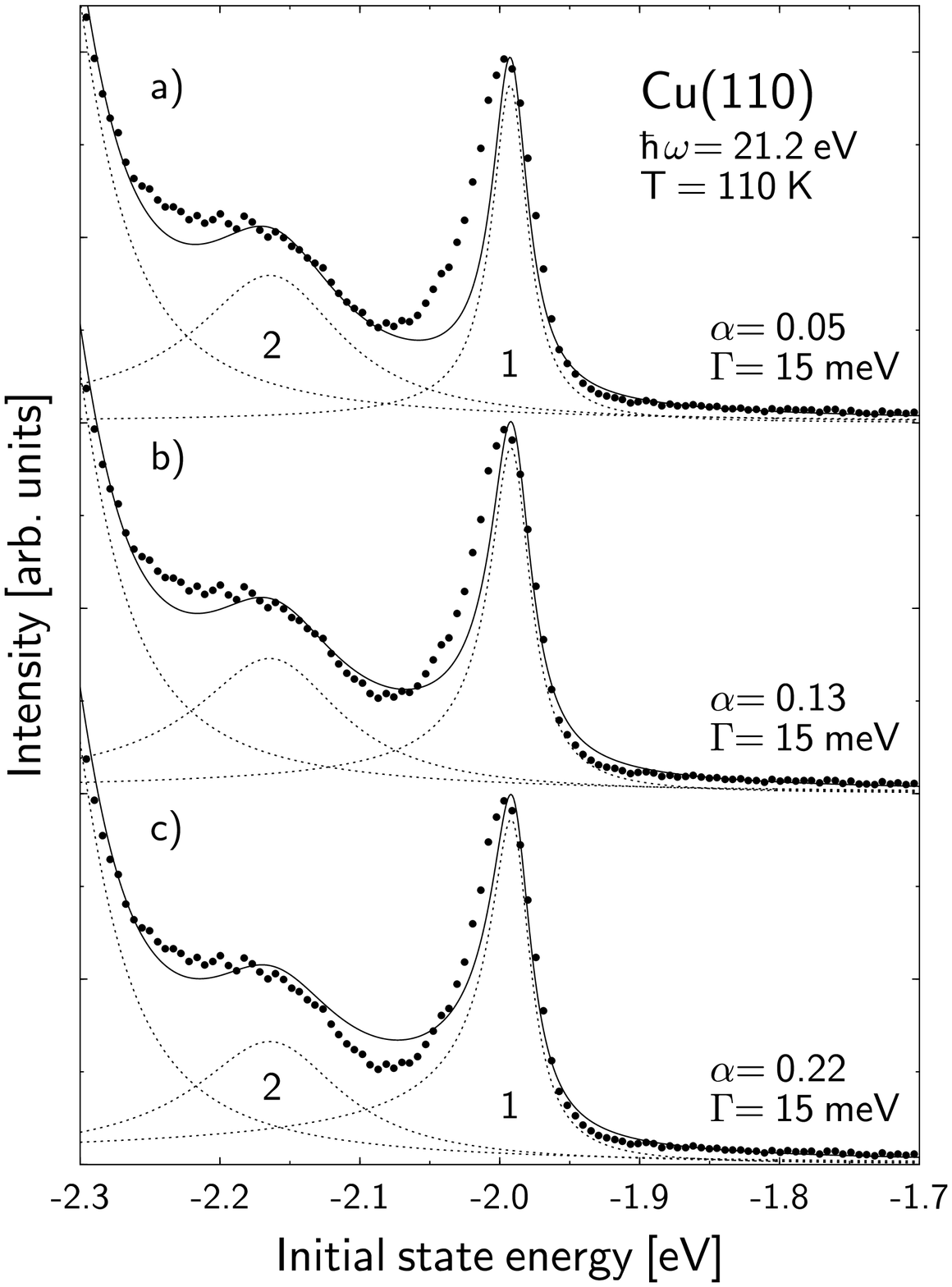}    
    \caption{Decomposition of the Cu(110) normal emission spectrum 
      obtained with $21.2 \, \mathrm{eV}$ using the Doniach-Sunjic
      lineshape with a half-width at half-maximum (HWHM) $\Gamma$ and
      an asymmetry parameter $\alpha$. The Shirley-type background has
      already been substracted from the data points}
    \label{fig:donsun}
  \end{center}
\end{figure}

As mentioned already peak 1 can not be described by a Lorentzian line
shape. The steep increase at the low binding energy side and its
clearly visible asymmetry on the opposite wing resembles the
Doniach-Sunjic lineshape\cite{doniach70} which, as is well known for
core-level excitation, has a high binding energy tail due to
electron-hole pair excitations around $E_F$. At the X-point the
$d$-band is flat and one might expect some response of the Fermi sea
to the rather well-localized photo-hole.  Fig.~\ref{fig:donsun} shows
three different lineshape calculations, which are based on the
Doniach-Sunjic formula\cite{wertheim78,doniach70}. The shape of peak 1
cannot be described satisfactorily, if we use an asymmetry parameter
$\alpha = 0.05$ with is typically observed in XPS spectra from copper
or silver\cite{wertheim78}, see the solid line in
Fig.~\ref{fig:donsun}(a).  Although the fit may be improved with
$\alpha \geq 0.1$, see Fig.~\ref{fig:donsun}(b) and
Fig.~\ref{fig:donsun}(c), one can argue whether these values describe
the response function of the $d$-hole. Actually we presume the degree
of delocalization to be stronger for the $3d$-bands than for core
holes and rule out an exclusive interpretation of the asymmetry due to
the Doniach-Sunjic excitation.

Instead we use line-shape modeling going beyond Lorentzians. First,
Lorentzians result provided both initial and final state bands
disperse linearly across the direct emission photoelectron
peak\cite{smith93}.  This is clearly violated at the X-point. Second,
$\Gamma_h$ ist regarded as constant on the energy scale of the
linewidth. The data collected in table~\ref{tab:linewidth} clearly
indicate a rapid variation of $\Gamma_h$ around the upper edge of the
$d$-band.  Therefore we have performed a numerical line shape
calculation following the ideas outlined in detail in subsection 3.2
to 3.4 of Ref.\ \onlinecite{matzdorf98}. The initial state bands are
taken from the calculation of Ref.\ \onlinecite{eckardt84}. The final
state is modeled by a free-electron-like band according to
$E_f(k_\perp) = V_0 + \hbar^2 k_\perp^2 /2m$ where $m$ is the free
electron rest mass and $V_0$ is chosen according to the experimentally
observed X-point.

\begin{figure}
  \begin{center}
    \includegraphics[width=14cm]{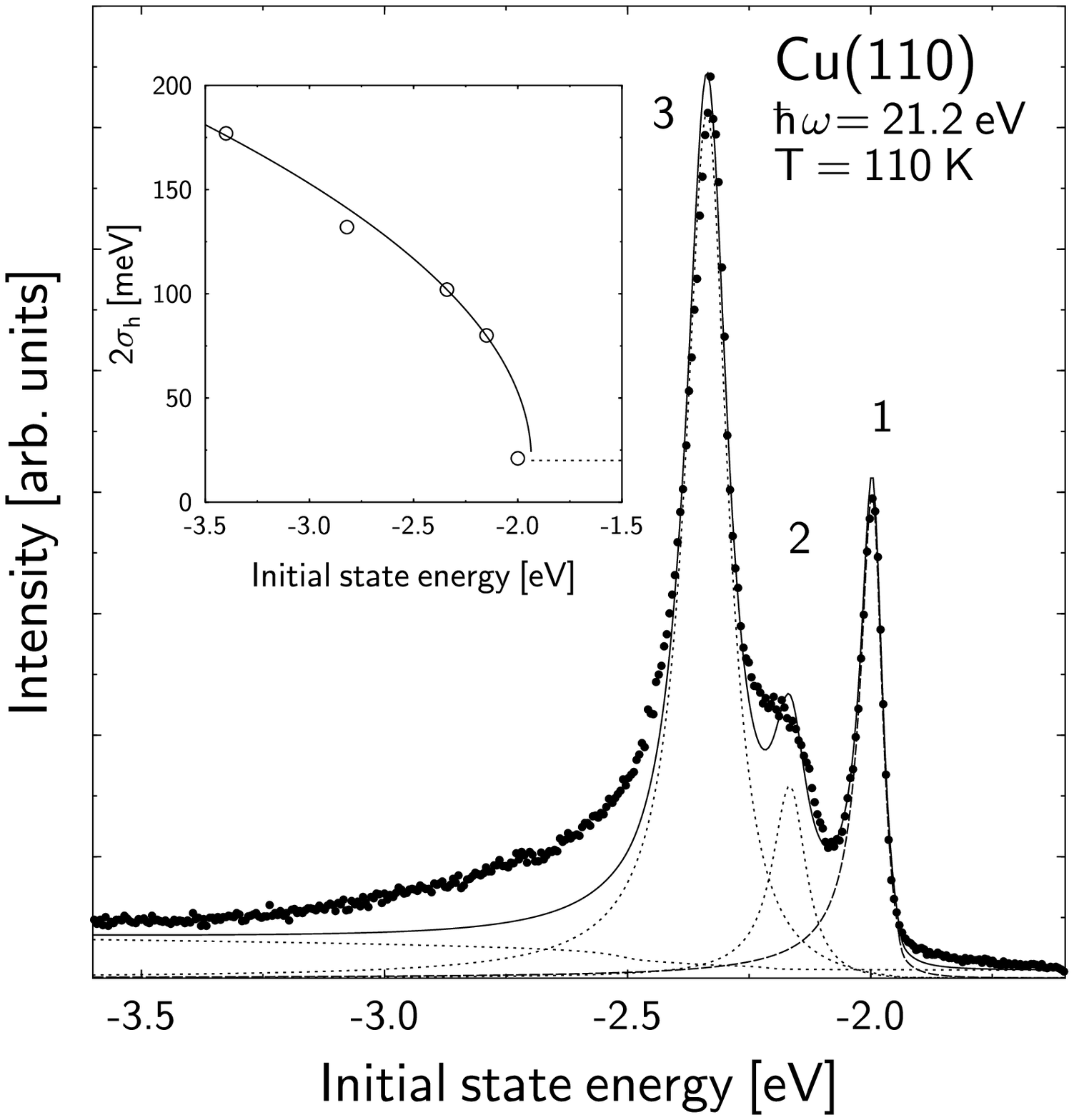}     
    \caption{Line shape analysis for a Cu(110) normal emission spectra 
      excited with $\hbar \omega = 21.2 \, \mathrm{eV}$ following the
      ideas outlined in the text. The lifetime parameterization
      $\sigma_h(E_i)$ shown in the inset is inspired by the measured
      linewidths (open circles) within the $d$-bands.}
    \label{fig:upsint}
  \end{center}
\end{figure}

The line shape is then given for each contributing transition by the
convolution (eq.~(20) of Ref.~\onlinecite{matzdorf98})
\begin{equation} 
  \label{eq:model}
  I(E_i) \propto \int dk_\perp \, \mathcal{L}_h (E_f-E_i-
  \hbar\omega , \sigma_h)\mathcal{L}_e(k_\perp^0-k_\perp-
  G_\perp , \sigma_e).
\end{equation} 
In eq.~(\ref{eq:model}) $\mathcal{L}_h$ and $\mathcal{L}_e$ are
Lorentzian distributions, whereas $\sigma_h$ and $\sigma_e$ represent
the corresponding half-widths at half-maximum. They are treated as
parameters in order to reproduce the experimental line shape. After
performing the integration given by equ.~(\ref{eq:model}) the resulting
$I(E_i)$ is additionally convoluted with a Gaussian to account for the
experimental energy resolution before comparison with the measured
line shape.  To model $\sigma_h$ we have chosen
\begin{equation} 
  \label{eq:sigma_h} 
  \sigma_h(E_i) = 64 \left(|E_i|-1.95\right)^{1/2} + 10\  \textnormal{[meV]}
\end{equation} 
for the energy range $E_i < -1.95\,$eV. Above the upper $d$-band edge,
i.e.\ $E_i > -1.95\,$eV, we simply assume $\sigma_h = 10\,$meV. Our
choice of 2$\sigma_h = \Gamma_h$  assumes that $\Gamma_h$ does not change
significantly with $k_\perp$ in the vicinity of the X-point. This
parameterization, however, implies a drastic decrease of $\Gamma_h$ on
approaching the $d$-band edge. For the corresponding final states we
use
\begin{equation} 
  \label{eq:sigma_e}
  \sigma_e (E_f) = \frac{dk_\perp}{dE_f}\  \Gamma_e \ ,
\end{equation}
where $E_f$ is the free-electron-like parabola mentioned above. The
linear relation $\Gamma_e = a (E_f-E_F)$ used in our calculation is an
empirical average over many photoemission results, see Ref.\ 
\onlinecite{goldmann91} for experimental data and Ref.\ 
\onlinecite{echenique00} for some theoretical background.  A typical
result of the numerically modeled peak shapes together with the
parameterization of eq.~(\ref{eq:sigma_h}) is reproduced in
Fig.~\ref{fig:upsint}. Obviously the agreement is very much improved
compared to simple Lorentzians and we believe the resulting data for
$\Gamma_h(E_i)$ around the X-point to be reliable.  At the upper
$d$-band edge we deduce $\Gamma_h = (20 \pm 5) \, \mathrm{meV}$.

On the basis of the experimental linewidth as given in
table~\ref{tab:linewidth} we calculate the corresponding hole
lifetimes $\tau_h = \hbar/\Gamma_h$, which are reproduced in
Fig.~\ref{fig:dholes2}.  The lifetime is very long at the top of the
$d$-band ($\tau_h = 31 \, \mathrm{fs}$), then decreases rapidly with
increasing distance from $E_F$ and seems to saturate finally.  This
behaviour is clearly not consistent with an energy dependence $\tau_h
\propto (E - E_F)^{-2}$.  In earlier
publications\cite{matzdorf94,matzdorf99} we had collected experimental
data taken from spectra measured with laboratory light sources, i.e.\ 
without the possibility to tune the photon energy exactly to the
symmetry points.  These results still seemed to follow the $(E -
E_F)^{-2}$-dependence.  The corresponding claim e.g.\ in Ref.\ 
\onlinecite{matzdorf99} is obviously ruled out by the new data.  In
fact it was Petek and his group\cite{petek99} who first found the
surprisingly long hole lifetime of $\tau_h \geq 24\, \mathrm{fs}$
($\Gamma_h = 28\, \mathrm{meV}$) at the upper $d$-band edge, and they
also argued that Fermi-liquid-like energy dependence should not hold
for the noble metal $d$-holes.

\section{Theoretical model and discussion}

For many years the theoretical framework of the e-e inelastic
lifetimes and mean free paths of excited electrons in metals had been
based on the free-electron gas (FEG) model of the solid.  In this
simple model and for energies very near the Fermi level $(|E-E_F| \ll
E_F)$, the inelastic lifetime is found to be, in the high-density
limit ($r_s\to 0$)\cite{rs}, $\tau(E) = 263 \, r_s^{-5/2} (E-E_F)^{-2}
\, \mathrm{fs}$, where $E$ and $E_F$ are expressed in
$\mathrm{eV}$\cite{quinn58}. Deviations from this simple formula,
which have been shown to be mainly due to band-structure
effects\cite{campillo00,schoene00}, were clearly
observed\cite{pawlik97,knoesel98,petek00}.  First-principles
caculations of the hole dynamics in the noble metals copper and gold
have been reported very recently\cite{campillopress}, too. Here we
focus on the calculation of $d$-hole lifetimes in copper and a
comparison with the experiment.

The basic quantity in the investigation of quasiparticle dynamics is
the probability per unit time for the probe quasiparticle (electron or
hole) to be scattered from a given initial state $\phi_{n, \mathbf{k}}
(\mathbf{r})$ of energy $E_{n, \mathbf{k}}$. In the framework of
many-body theory, this probability is identified with the inverse
quasiparticle lifetime (we use atomic units, i.e., $e^2=\hbar=m_e=1$),
\begin{equation}
  \label{eq:lifetime} 
  \tau^{-1}=-2\int d \mathbf{r}\int d \mathbf{r}'\,\phi^*_{n,\mathbf{k}}
  ( \mathbf{r})\,\mathrm{Im}\Sigma(\mathbf{r},\mathbf{r}';E_{n,\mathbf{k}})\,
  \phi_{n,\mathbf{k}}(\mathbf{r}'),
\end{equation}
where $\Sigma(\mathbf{r},\mathbf{r}';E_{n,\mathbf{k}})$ is the
so-called self-energy of the quasiparticle.  In the GW approximation,
one keeps only the first-order term in a series expansion of the
self-energy in terms of the screened interaction.  For details on
further approximations we refer to Ref.~\onlinecite{echenique00}.

\begin{figure}
  \begin{center}
    \includegraphics[width=14cm]{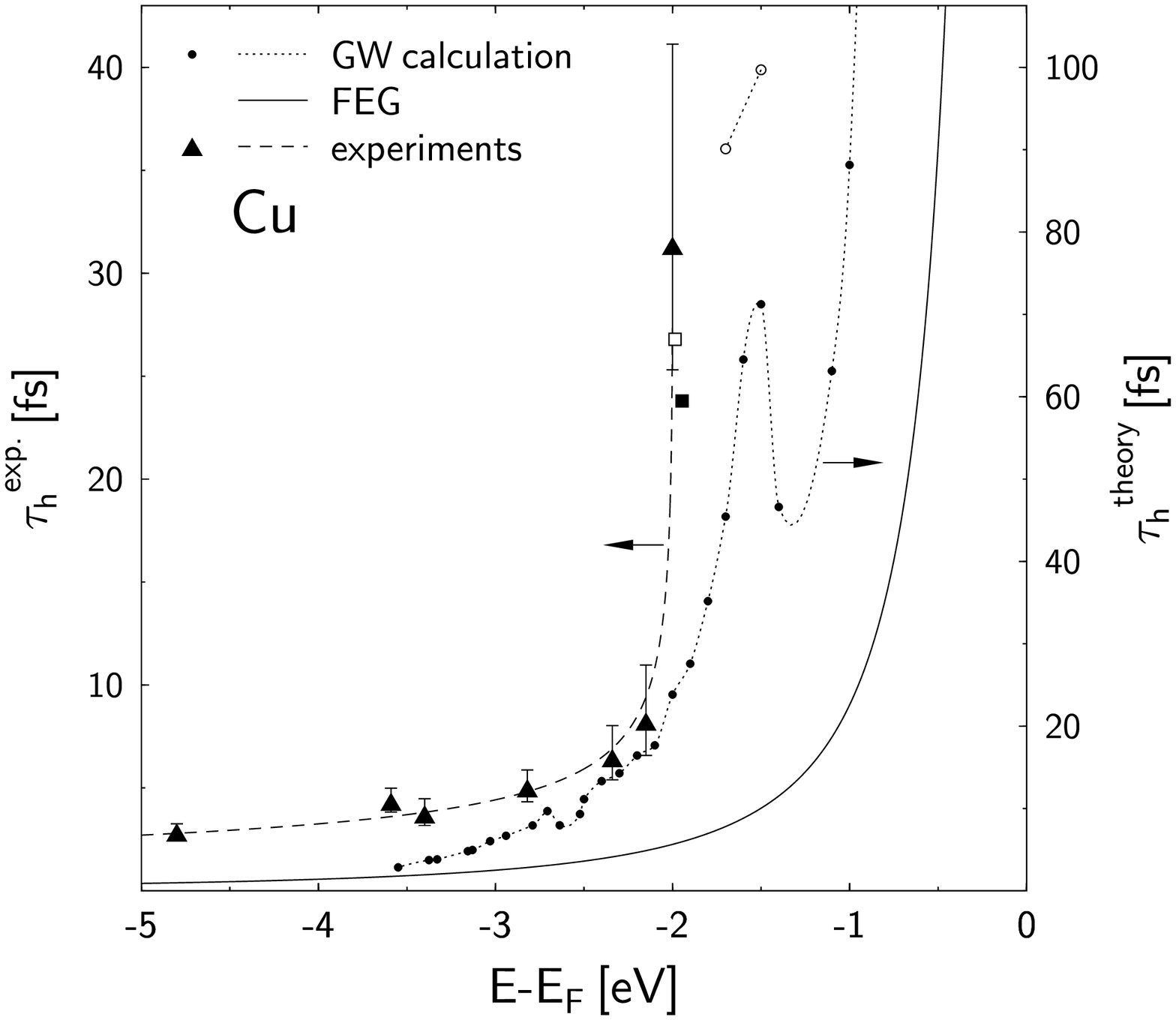}
    \caption{Comparison of lifetimes $\tau_h = \hbar /
      \Gamma_h$ in copper due to electron-hole interaction. Triangles
      represent the experimental data according to
      table~\ref{tab:linewidth} obtained for $d$-holes at the
      $\Gamma$- and X-point of the Brillouin zone. Filled (open)
      circles represent hole lifetimes obtained from the full
      band-structure calculation with (without) averaging over all
      wave vectors and bands.  The solid line shows the corresponding
      FEG calculation with $r_s=2.67$.  Note the different ordinate
      scales for experimental (left) and theoretical (right) results.
      Experimental data from other authors: filled square Ref.\ 
      \protect\onlinecite{purdie98}\protect, open  square Ref.\ 
      \protect\onlinecite{petek99}\protect}
    \label{fig:dholes2}
  \end{center}
\end{figure}

Our full band-structure calculations of hole lifetimes in copper, as
obtained from eq.~(\ref{eq:lifetime}) by averaging over all wave
vectors and bands with the same energy, are shown in
Fig.~\ref{fig:dholes2} (solid circles), together with the FEG
calculations with $r_s=2.67$.  There is a qualitatively good agreement
of our theoretical calculations and the experimental data, which are
both well above the FEG prediction.  Band-structure effects, which
significantly enhance the lifetimes predicted within a FEG model of
the solid, are found to be mainly due to a major contribution from
occupied $d$-states participating in the screening of e-e
interactions, and also to the small overlap of the initial and final
$d$-hole and $sp$-states below the Fermi level\cite{note0}.

Density-functional theory\cite{gross96} (DFT) predicts an upper
$d$-band edge that is located $\sim 1.5\, \mathrm{eV}$ below the Fermi
level, i.e.\ $\sim 0.5\, \mathrm{eV}$ higher than observed by
photoemission experiments\cite{huefner95,strocov98}. Since the wave
functions and energies entering eq.~(\ref{eq:lifetime}) have been
obtained by DFT, the calculation yields enhanced lifetimes at $E - E_F
= -1.5 \, \mathrm{eV}$\cite{note1}, whereas the measured lifetimes
increase near the actual upper $d$-band edge at $2.0\, \mathrm{eV}$
below $E_F$, see Fig.~\ref{fig:dholes2}.  As the energy $|E-E_F|$
increases the larger phase space for the hole decay yields shorter
lifetimes, but the small overlap of $d$- and $sp$-states still
prevents the establishment of a free-electron-like behaviour $\tau
\propto(E-E_F)^{-2}$.  While our calculated lifetimes below $\sim 3 \,
\mathrm{eV}$ tend to approach those predicted by a FEG model of the
solid, the experimental data show a distinct asymptotic behaviour at
the large binding energies $|E-E_F|$. This discrepancy between theory
and experiment may be, again, due to the failure of the DFT to
reproduce the actual band structure of copper.  Indeed, experiments
and an accurate mapping of the occupied band structure\cite{strocov98}
have shown that not only the actual upper $d$-band edge is located
$\sim 0.5\, \mathrm{eV}$ lower than predicted by DFT, but also the
$d$-band width is $2.7$ ($1.0$) eV at the X- ($\Gamma$-) point, i.e.
$\sim 0.3$ ($0.13$) eV narrower than in the case of the DFT band
structure. This narrowing of the $d$-band points to a stronger
decoupling between $d$- and $sp$-states, as well as a localization and
hybridization of the states below the Fermi level.  Thus the hole
lifetimes at the upper $d$-band edge are expected to be even longer
than predicted by our calculations.  Hence, the departure of the
quasiparticle wave-functions from their DFT counterparts is expected
to prevent the establishment of a free-electron-like behavior, thereby
providing a qualitative explanation for the long lifetimes measured
below $-3\, \mathrm{eV}$ (see Fig.~\ref{fig:dholes2}).

Comparing the experimental and theoretical data we have to keep in
mind that both data sets are not completely compatible: First, theory
integrates over the full Brillouin zone, whereas the experimental data
refer to selected $k$-space points at X and $\Gamma$. Second, due to
the shifted $d$-band edge in the DFT calculation the available phase
space for Auger recombination is underestimated, yielding calculated
lifetimes which are too long. Third, $sp$-like bands contribute to the
calculated data, in particular above the the $d$-band threshold.
Finally, the band structure calculation is performed
non-relativistically, with the symmetry admixtures of different
orbital character neglected and splittings due to the spin-orbit
coupling at X and $\Gamma$ omitted. Regarding these deficiencies the
basic trends are nevertheless well reproduced. There is a threshold
like onset of large $d$-hole lifetimes at the upper $d$-band edge,
both in experiment and theory.

\section{Summary and conclusions}

We have presented high-resolution angle-resolved photoemission
experiments and many-body quasiparticle GW calculations of e-e
inelastic lifetimes of $d$-holes in copper. Both our theoretical
calculations and the experimental measurements have been found to be
well above the FEG predictions, showing the important role that
band-structure effects play in the hole-decay mechanism. The
$(E-E_F)^{-2}$-dependence of the hole lifetimes is based on the
free-electron picture. Flat bands, as occuring near the symmetry
points of noble metals, can lead to rather different energy
dependence. The experimentally observed decrease of the $d$-hole
lifetimes by approximately one order of magnitude within the $d$-bands
is attributed to the large phase space and the effective $d$-$d$
scattering. Since the $d$-holes are much more confined to the ionic
core, the $d$-$d$ scattering cross section should be much larger than
the $d$-$sp$ cross section.


Although there is qualitative good agreement of theory and experiment,
there still exist significant discrepancies. On the one hand, both
theory and experiment exhibit the longest lifetimes at the top of the
$d$-bands, due to the small overlap of $d$- and $sp$-states below the
Fermi level. Differences in the position of the maximum lifetime are
attributed to the failure of the DFT band-structure calculation. On
the other hand the distinct behaviour of the lifetimes at large
energies $|E-E_F|$ suggest that the narrowing of the $d$-band width
may play an important role, thereby showing the need to go beyond DFT
in the description of the band structure of copper.

\begin{acknowledgements}
  We acknowledge support by the University of the Basque Country, the
  Basque Hezkuntza, Unibertsitate eta Ikerketa Saila, the Spanish
  Ministerio de Educaci\'on y Cultura, and the European Union Research
  Training Network program NANOPHASE.  P.M.E. gratefully acknowledges
  support from the Max Planck Research Award funds. Our experimental
  work is continuously supported by the Deutsche
  Forschungsgemeinschaft (DFG).  The experiments using synchrotron
  radiation are financed by the German Bundesministerium f\"ur Bildung
  und Forschung (BMBF). We are also indebted to the referees of the
  first draft of the manuscript, whose extensive comments helped to
  improve this paper considerably.
\end{acknowledgements}


\begin{thebibliography}{99} 
  
\bibitem{dai95} H. L. Dai and W. Ho (Eds.), \textit{Laser Spectroscopy
    and Photochemistry on Metal Surfaces} (World Scientific,
  Singapore, 1995).
  
\bibitem{petek97} H. Petek and S. Ogawa, Progr. Surf. Sci.
  \textbf{56}, 239 (1997).
  
\bibitem{petek00} H. Petek and A. F. Heinz (Eds.), Special Issue on
  \textit{Electron Dynamics in Metals}, Chem. Phys. \textbf{251},
  1-329 (2000).
  
\bibitem{matzdorf98} R. Matzdorf, Surf. Sci. Rep. \textbf{30}, 153
  (1998).
  
\bibitem{theilmann97} F. Theilmann, R. Matzdorf, G.  Meister, and A.
  Goldmann, Phys. Rev.  B \textbf{56}, 3632 (1997).
  
\bibitem{pawlik97} S. Pawlik, M. Bauer, and M. Aeschlimann, Surf. Sci.
  \textbf{377}, 206 (1997).
  
\bibitem{knoesel98} E. Knoesel, A. Hotzel, and M. Wolf, Phys. Rev. B
  \textbf{57}, 12812 (1998).
  
\bibitem{petek00.2} H. Petek, H.  Nagano, M.J. Weida, and S. Ogawa,
  Chem. Phys. \textbf{251}, 71 (2000).
  
\bibitem{echenique00} P.M. Echenique, J.M. Pitarke, E.V. Chulkov, and
  A. Rubio, Chem. Phys.  \textbf{251}, 1 (2000). This review contains
  many references to earlier work.
  
\bibitem{buergi99} L. B\"urgi, O. Jeandupeux, H. Brune, and K. Kern,
  Phys.  Rev. Lett.  \textbf{ 82}, 4516 (1999).
  
\bibitem{matzdorf99} R. Matzdorf, A. Gerlach, F. Theilmann, G.
  Meister, and A.  Goldmann, Appl. Phys. B \textbf{68}, 393 (1999).
  
\bibitem{kevan92} S. D. Kevan (Ed.), \textit{Angle-resolved
    Photoemission, Studies in Surface Science and Catalysis}, Vol. 74
  (Elsevier, Amsterdam, 1992).
  
\bibitem{huefner95} S. H\"ufner, \textit{Photoelectron
    Spectroscopy-Principles and Applications}, Springer Series in
  Solid-State Physics, Vol. 82 (Springer, Berlin, 1995).

\bibitem{smith93} N. V.  Smith, P. Thiry, and Y. Petroff, Phys. Rev. B
  \textbf{47}, 15476 (1993).
  
\bibitem{mcdougall95} B.A.  McDougall, T. Balasubramanian, and E.
  Jensen, Phys.  Rev. B \textbf{51}, 13891 (1995).

\bibitem{campillo00} I. Campillo, J. M. Pitarke, A. Rubio, E. Zarate,
  and P.  M.  Echenique, Phys. Rev. Lett. \textbf{83}, 2230 (1999); I.
  Campillo, V.  M. Silkin, J.  M. Pitarke, E. V. Chulkov, A. Rubio,
  and P. M.  Echenique, Phys. Rev. B \textbf{61}, 13484 (2000); I.
  Campillo, J. M.  Pitarke, A. Rubio, and P. M. Echenique, Phys.  Rev.
  B \textbf{62}, 1500 (2000).
  
\bibitem{schoene00} W.-D. Sch\"one, R. Keyling, and W. Ekardt, Phys.
  Rev. B \textbf{60}, 8616 (1999); R. Keyling, W.-D. Sch\"one, and W.
  Ekardt, Phys. Rev. B \textbf{61}, 1670 (2000).
  
\bibitem{campillopress} I.  Campillo, A. Rubio, J. M. Pitarke, A.
  Goldmann, and P. M.  Echenique, Phys. Rev. Lett. \textbf{85}, 3241
  (2000)
  
\bibitem{janowitz99} C. Janowitz, R. M\"uller, T. Plake, Th. B\"oker,
  and R.  Manzke, J.  Electron Spectrosc. Rel. Phen. \textbf{105}, 43
  (1999).
  
\bibitem{courths84} R. Courths and S. H\"ufner, Phys. Rep.
  \textbf{112}, 53 (1984).
  
\bibitem{shirley72} D. A. Shirley, Phys. Rev. B \textbf{5}, 4709 (1972)

\bibitem{dietz79} E. Dietz and F.J. Himpsel, Solid State Commun.
  \textbf{30}, 235 (1979).

\bibitem{matzdorf94} R. Matzdorf, R. Paniago, G. Meister, and A.
  Goldmann, Solid State Commun. \textbf{92}, 839 (1994).

\bibitem{matzdorf96} R.  Matzdorf, G. Meister, and A.  Goldmann, Phys.
  Rev. B \textbf{54}, 14807 (1996).
    
\bibitem{kevan83} S. D.  Kevan, Phys. Rev. Lett. \textbf{50}, 526
  (1983).
  
\bibitem{tersoff83} J. Tersoff and S.D. Kevan, Phys. Rev. B
  \textbf{28}, 4267 (1983).

\bibitem{eckardt84} H. Eckardt, L. Fritsche, J. Noffke, J. Phys. F
  \textbf{14}, 97 (1984)
    
\bibitem{purdie98} D. Purdie, M. Hengsberger, M. Garnier, and Y. Baer,
  Surf.  Sci. \textbf{407}, L671 (1998).

\bibitem{petek99} H. Petek, H. Nagano, and S. Ogawa, Phys. Rev.  Lett.
  \textbf{83}, 832 (1999).

\bibitem{doniach70} S. Doniach, M. Sunjic, J. Phys. C \textbf{3}, 285
  (1970)
  
\bibitem{wertheim78} G. K. Wertheim, P. H. Citrin in
  \textit{Photoemission in Solids I}, edited by M. Cardona and L. Ley,
  Topics in Appl. Phys., Vol. 26, (Springer Verlag, Berlin, 1978)

\bibitem{goldmann91} A. Goldmann, W. Altmann, V. Dose, Solid State
  Comm. \textbf{79}, 511 (1991)

\bibitem{rs} The parameter $r_s$ is defined by the relation $1/n=4
  \pi(r_sa_0)^3/3$, $n$ and $a_0$ being the average electron density
  and the Bohr radius, respectively.
  
\bibitem{quinn58} J. J. Quinn and R. A. Ferrell, Phys. Rev.
  \textbf{112}, 812 (1958).
  
\bibitem{note0} The contribution from occupied $d$-states
  participating in the screening of e-e interactions is also present
  in the hot-electron decay.  However, the low overlap of the hole
  initial $d$- and final $sp$-states is absent in the case of excited
  electrons above the Fermi level, which explains the fact that holes
  exhibit longer lifetimes than electrons with the same excitation
  energy.

\bibitem{gross96}E. K. U. Gross, F. J. Dobson, and M. Petersilka,
  \textit{Density Functional Theory} (Springer, New York, 1996).
  
\bibitem{strocov98} V. N. Strocov, R.  Claessen, G. Nicolay, S.
  H\"ufner, A.  Kimura, A.  Harasawa, S. Shin, A. Kakizaki, P. O.
  Nilsson, H. I.  Starnberg, and P. Blaha, Phys.  Rev. Lett.
  \textbf{81}, 4943 (1998).  Our first-principles calculations predict
  significant deviations of the unoccupied bands from free-electron
  behaviour, in agreement with these experiments.
  
\bibitem{note1} At the top of our calculated DFT $d$-bands, at the
  X-point with $E=-1.5\, \mathrm{eV}$ and $E=-1.7\, \mathrm{eV}$, we
  obtain $d$-hole lifetimes of $100$ and $90\, \mathrm{fs}$,
  respectively. These are included as open circles in
  Fig.~\ref{fig:dholes2}.
                
\end{thebibliography}
\end{document}